\documentclass[english,letterpaper,aps,prl,twocolumn,showpacs,superscriptaddress]{revtex4-1}
\usepackage{lmodern}

\usepackage[T1]{fontenc}
\usepackage[latin1]{inputenc}
\usepackage{graphicx}
\usepackage{amssymb}

\makeatletter

\usepackage{lmodern}
\usepackage{subfig}

\makeatletter

\usepackage{lmodern}

\makeatletter

\usepackage{lmodern}

\makeatletter

\usepackage{lmodern}

\makeatletter

\makeatletter

\usepackage{epsfig}

\usepackage{dcolumn}

\usepackage{bm}

\usepackage{bbm}

\usepackage{wasysym}

\newlength{\textwidthm}

\makeatother

\makeatother

\makeatother

\makeatother

\makeatother

\usepackage{babel}
\makeatother
\begin{document}

\title{Space Group Symmetry Classification of Energy Bands in Graphene
and Destruction of Dirac Points by Supercell Potential}

\author{E. Kogan}
\email{Eugene.Kogan@biu.ac.il}
\affiliation{Department of Physics, Bar-Ilan University, Ramat-Gan 52900,
Israel}
\affiliation{Departamento de Fisica de Materiales, Facultad de Ciencias Quiimicas,
Universidad del Pais Vasco/Euskal Herriko Unibertsitatea, DIPC,
San Sebastian/Donostia, 20080 Basque Country, Spain}
\author{V. U. Nazarov}
\email{nazarov@gate.sinica.edu.tw}
\affiliation{Research Center for Applied Sciences, Academia Sinica, Taipei 11529, Taiwan}
\author{V. M. Silkin}
\email{vycheslav.silkin@ehu.es}
\affiliation{Departamento de Fisica de Materiales, Facultad de Ciencias Quiimicas,
Universidad del Pais Vasco/Euskal Herriko Unibertsitatea, DIPC,
San Sebastian/Donostia, 20080 Basque Country, Spain}
\author{E. E. Krasovskii}
\email{eugene.krasovskii@ehu.es}
\affiliation{Departamento de Fisica de Materiales, Facultad de Ciencias Quiimicas,
Universidad del Pais Vasco/Euskal Herriko Unibertsitatea, DIPC,
San Sebastian/Donostia, 20080 Basque Country, Spain}
\date{\today}

\begin{abstract}
In the previous publications (E. Kogan and V. U. Nazarov, Phys. Rev. B {\bf 85}, 115418 (2012) and E. Kogan, Graphene {\bf 2}, 74 (2013))
we presented point group symmetry classification of energy bands in graphene. In the present note we
generalize classification by considering the space group symmetry.
Also in the framework of group theory we  describe destruction of the Dirac points in graphene by a supercell potential.
\end{abstract}

\pacs{73.22.Pr}

\maketitle

\section{Introduction}

Understanding of the symmetries of the electrons dispersion law in graphene is of crucial importance. Actually, the symmetry classification of the energy bands in graphene (or "two-dimensional graphite") was presented nearly 60 years ago by Lomer in his seminal paper \cite{lomer}. Later the subject was analyzed by Slonczewski and Weiss \cite{slon}, Dresselhaus and Dresselhaus \cite{dresselhaus}, Bassani and Parravicini \cite{bassani}. Recent approaches to the problem are presented in the papers by Malard et al. \cite{malard}, Manes \cite{manes}.
Different  approaches to the symmetry classification  are based on different methods of applications of group theory.

In our paper \cite{kogan1} the labeling of the bands was based on compatibility relations and guesses.
 In the next publication \cite{kogan2}
we have shown that in the framework of the tight-binding approximation the representations of the little group in the symmetry points can be rigorously found in the framework of the group theory algebra. Though the idea of using the tight-binding approximation is by no means new (it was used already in the work by Lomer), our approach is  different,  and, to our opinion, more convenient for applications.

Unfortunately, however well the subject is studied, there is always an opportunity to make a mistake. The mistake in our symmetry classifications
of the bands at the point $\Gamma$ was pointed in the recent paper \cite{elder}. Motivated by that criticism, we reanalyzed the subject, found the source of the mistake
and corrected it. This analysis, which is intended to be read together with Ref. \cite{kogan2}, we start the paper from.
Then we present classification of the electron bands in the most interesting points (the corners of the Brillouine zone)
considering space symmetry group. Finally we show how the symmetry of the states changes under application of supercell potential.

\section{Tight--binding model}

Our tight-binding model space includes four atomic orbitals: $|s,p>$. (Notice that we assume only symmetry of the basis functions with respect to rotations and reflections; the question how these functions are connected with the atomic functions of the isolated carbon atom is irrelevant.)
We search for the solution of Schroedinger equation as a linear combination of the functions
\begin{eqnarray}
\label{tb}
\psi_{\beta;{\bf k}}^j=\sum_{{\bf R}_j}\psi_{\beta}\left({\bf r}-{\bf R}_j\right) e^{i{\bf k\cdot R}_j},
\end{eqnarray}
where $\psi_\beta$ are atomic orbitals, $j=A,B$ labels the sub-lattices, and  ${\bf R}_j$ is the radius vector of an atom in the sublattice $j$.
A  symmetry transformation of the functions $ \psi_{\beta;{\bf k}}^j$ is a direct product of two transformations: the transformation of the sub-lattice functions $\phi^{A,B}_{{\bf k}}$, where
\begin{eqnarray}
\label{2}
\phi_{\bf k}^j=\sum_{{\bf R}_j} e^{i{\bf k\cdot R}_j},
\end{eqnarray}
and the transformation of the orbitals $\psi_{\beta}$. Thus the representations realized by the functions (\ref{tb}) will be the direct product of two representations.

The Hamiltonian of graphene being symmetric with respect to reflection in the graphene plane, the bands built from the $|z>$  orbitals decouple from those built from the $|s,x,y>$  orbitals. The former are odd with respect to reflection, the latter are even. In other words, the former form $\pi$  bands, and the latter form $\sigma$  bands.

The group of wave vector ${\bf k}$  at the point $\Gamma$ is $D_{6h}$, at the point $K$  is $D_{3h}$, at the point $M$ is $D_{2h}$, at the lines constituting triangle $\Gamma-K-M$  is $C_{2v}$  \cite{thomsen}.
The  representations of the group  $D_{6h}$ can be obtained on the basis of the identity
\begin{eqnarray}
\label{identity}
D_{6h}= C_{6v}\times C_s.
\end{eqnarray}

We have found \cite{kogan2} that   at the point $\Gamma$  the orbitals $|z>$ realize
\begin{eqnarray}
\label{a}
(A_{1}+B_2)\times A''
\end{eqnarray}
representation, the orbitals $|s>$ -
\begin{eqnarray}
(A_{1}+B_2)\times A'
\end{eqnarray}
representation,
and the orbitals $|x,y>$ -
\begin{eqnarray}
\label{c}
(E_{1}+E_{2})\times A'
\end{eqnarray}
representation of the group $D_{6h}$. In Eqs. (\ref{a}) - (\ref{c}) the first multiplier refers to the irreducible representations of the group $C_{6v}$, and the second multiplier refers to the irreducible representations of the group $C_s$ (the character tables are presented in Table \ref{table:d2}).

\begin{table}
\begin{tabular}{|l|l|rr|}
\hline
$C_s$ & & $E$ &  $\sigma$ \\
& $C_i $ & $E$ &  $I$ \\
\hline
$A'$ & $Ag$ & 1 & 1  \\   $A''$ & $A_u$ &1  & $-1$     \\
\hline
\end{tabular}
\begin{tabular}{|l|l|l|rrrrrr|}
\hline
$C_{6v}$ & &  & $E$ & $C_2$ & $2C_3$ & $2C_6$ & $3\sigma_v$ & $3\sigma_v'$ \\
& $D_6$ &  & $E$ & $C_2$ & $2C_3$ & $2C_6$ & $3U_2$ & $3U_2'$ \\
& & $D_{3h}$   & $E$ & $\sigma$ & $2C_3$ & $2S_3$ & $3U_2$ & $3\sigma_v$ \\\hline
$A_{1}$ & $A_{1}$ & $A_1'$ & 1 & 1 & 1 & 1 & 1 & 1 \\
$A_{2}$ & $A_{2}$ & $A_2'$ & 1 & 1 & 1 & 1 & $-1$ & $-1$ \\
$B_{2}$  & $B_{1}$ & $A_1''$ & 1 & $-1$ & 1 & $-1$ & 1 & $-1$ \\
$B_{1}$  & $B_{2}$  &  $A_2''$ &1 & $-1$ & 1 & $-1$ & $-1$ & 1 \\
$E_{2}$  & $E_{2}$  & $E'$ & 2 & 2 & $-1$ & $-1$ & 0 & 0 \\
$E_{1}$  & $E_{1}$ & $E''$  & 2 & $-2$  & $-1$ & 1 & 0 & 0 \\
\hline
\end{tabular}
\caption{Character table for irreducible representations of  $C_s,C_i$  and $C_{6v},D_6,D_{3h}$ point groups}
\label{table:d2}
\end{table}

The irreducible representations of the group $D_{6h}$ are traditionally labelled not  on the basis of the identity   (\ref{identity}),
but on the basis of the alternative identity
\begin{eqnarray}
D_{6h}= D_{6}\times C_i.
\end{eqnarray}
Thus each representation of the group  $D_{6}$, say $A_1$, begets two representations: even $A_{1g}$  and odd $A_{1u}$.

To decompose the product of representations (\ref{a}) - (\ref{c}) with respect to the irreducible representations of the group $D_{6h}$ we need
to express the products of the symmetry operations of the groups $C_{6v}$ and $C_s$ through the products of the symmetry operations of the groups $D_{6}$ and $C_i$.
\begin{table}
\begin{tabular}{|c|c|c|c|c|c|c|c|c|c|c|c|}
\hline
$E$ & $C_2$ & $2C_3$ & $2C_6$ & $3U_2$ & $3U_2'$ & $I$ & $C_2I$ & $2C_3I$ & $2C_6I$ & $3U_2I$ & $3U_2'I$ \\
$E$ & $C_2$ & $2C_3$ & $2C_6$ & $3\sigma_v\sigma$ & $3\sigma_v'\sigma$   & $C_2\sigma$ & $\sigma$ & $2C_6\sigma$ & $2C_3\sigma$ & $3\sigma_v'$ & $3\sigma_v$ \\
\hline
\end{tabular}
\caption{Correspondence between the products of the symmetry operations of the groups $D_{6}$ and $C_i$ and the products of the symmetry operations of the groups $C_{6v}$ and $C_s$.}
\label{table:d222}
\end{table}
Using elementary algebra we obtain
\begin{eqnarray}
\begin{array}{l}
A_{1}\times A'=A_{1g}\\
B_{2}\times A'=B_{1u}\\
A_{1}\times A''=A_{2u}\\
B_{2}\times A''=B_{2g}\\
E_{1}\times A'=E_{1u}\\
E_{2}\times A'=E_{2g}\end{array}.
\end{eqnarray}

We have found \cite{kogan2} that   at the point $K$  the orbitals $|z>$ realize
$E''$
representation, the orbitals $|s>$ realize
$E'$
representation,
and the orbitals $|x,y>$ realize
$A_1'+A_2'+E'$ representation of the group $D_{3h}$.

\begin{figure}[h]
\begin{center}
\centering
\includegraphics[width=0.45\textwidth]{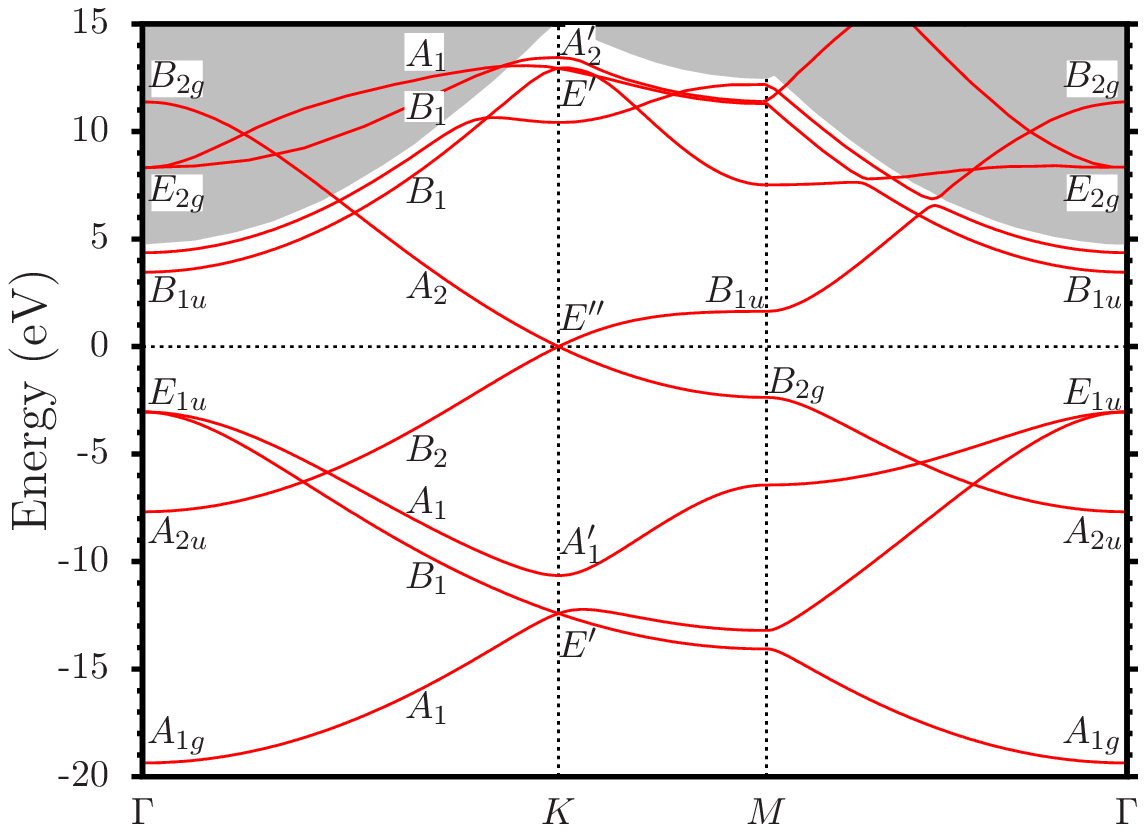}
\caption{\label{fig:bands}  (Color online) Graphene band structure evaluated with
use of the FP-LAPW method and the code Elk \cite{elk}. The dashed line shows the Fermi energy. }
\end{center}
\end{figure}

After that, using
the    compatibility relations presented in Table \ref{table:c} \cite{thomsen,kogan1}, we obtain  classification of the energy bands in graphene, presented on Fig. \ref{fig:bands}.
\begin{table}
\begin{tabular}{|c|c|c|}
	\hline
$C_{2v}$ & $D_{6h}$ & $D_{3h}$  \\
\hline
Rep &\multicolumn{2}{c|} {Compatible with}  \\
	\hline
$A_{1}$   & $A_{1g},B_{2u},E_{1u},E_{2g}$ & $A_1',E'$  \\
$A_{2}$   & $A_{1u},B_{2g},E_{1g},E_{2u}$ & $A_1'',E''$   \\
$B_{1}$   & $B_{1u},A_{2g},E_{1u},E_{2g}$ & $A_2',E'$  \\
$B_{2}$   & $A_{2u},B_{1g},E_{1g},E_{2u}$ & $A_2'',E''$  \\
	\hline
\end{tabular}
\caption{Compatibility relations}
\label{table:c}
\end{table}

In addition to what was done in our previous publications \cite{kogan1,kogan2}, we analyzed the symmetry of the $\pi$ bands at the point $M$. Irreducible representations of the point group $D_{2h}$ are obtained on the basis of identity
\begin{eqnarray}
D_{2h}= D_{2}\times C_i.
\end{eqnarray}
 \begin{table}
\begin{tabular}{|l|rrrr|}
\hline
 $D_2$ & $E$ & $C_2^z$ & $C_2^y$ &  $C_2^x$ \\
\hline
 $A$ & $1$ & 1 & 1 & 1 \\
 $B_3$ & $1$ & $-1$  & $-1$ & 1    \\
 $B_1$ & $1$ & 1 & $-1$ & $-1$ \\
 $B_2$ & $1$ & $-1$ & 1 & $-1$     \\
\hline
\end{tabular}
\caption{Character table for irreducible representations of   $D_2$ point group}
\label{table:d85}
\end{table}

Consider  the point ${\bf M}=\left(\frac{2\pi}{3a},0\right)$.
In the linear space spanned by the functions $\psi_{z;{\bf M}}^{A,B}$ we get $\chi(E)=2,\chi(C_2^x)=-2, \chi(C_2^yI)=2,\chi(C_2^zI)=-2$. All the other traces are equal to zero.
Hence  the functions $\psi_{z;{\bf M}}^{A,B}$ realize
\begin{eqnarray}
R=B_{1u}+B_{2g}
\end{eqnarray}
representation of the group.

It was recently shown \cite{Nazarov-13}
that parts of the 2D bands inside the 3D continuum (gray background in Fig. 1)
turn from true bound-state bands into scattering resonances, by acquiring a finite life-time due to the
coupling of the in-plane and perpendicular motions.

\section{Space group analysis}

In our previous publications \cite{kogan1,kogan2} classification
of the energy bands in graphene was done on the basis of point group analysis.
However, more general analysis can be performed on the basis of space group symmetry \cite{bradley,aroyo}.

Space group of graphene is $P6/mmm$.
Consider  the most interesting and practically important case of the symmetry analysis at the points ${\bf K},{\bf K}'$.
The little group at the point ${\bf K}$ (${\bf K}'$) is $P3/mm$.
Because $P3/mm$ is a symmorphic group, matrices of the allowed irreps of the little group $\overline{D}^{{\bf K},i}$ can be easily connected with the
matrices of the small representations $D^i$ of the little co-group
$D_{3h}$
\begin{eqnarray}
\label{generator2}
D^{{\bf K},i}(\tilde{W}, {\bf t})=e^{-i{\bf K}\cdot {\bf t}} D_{3h}^{i}(\tilde{W}).
\end{eqnarray}
In the l.h.s. of Eq. (\ref{generator2}) the first term in the brackets  is a point symmetry operator of the little co-group, and the second term is the translation vector
\begin{eqnarray}
\label{n}
{\bf t}=n_1{\bf t}_1+n_2{\bf t}_2=\frac{n_1}{2}(3,\sqrt{3})+\frac{n_2}{2}(3,-\sqrt{3}).
\end{eqnarray}

The generators of $D_{3h}$ are
$E,\sigma,C_3,U_2$. For one dimensional representations the matrices coincide with the traces in Table  \ref{table:d2}.  The matrices
for the two dimensional representations are presented in Table \ref{table:d7}.
 \begin{table}
\begin{tabular}{|l|cccc|}
\hline
$D_{3h}$ & E & $\sigma$ & $C_3$  & $U_2$ \\
\hline
$D^{E'}$ & $\left(\begin{array}{cc} 1 & 0 \\ 0 & 1  \end{array}\right)$ & $\left(\begin{array}{cc} 1 & 0 \\ 0 & 1  \end{array}\right)$ &  $\left(\begin{array}{cc} 0 & -1 \\ 1 & -1  \end{array}\right)$
   &  $\left(\begin{array}{cc} 0 & 1 \\ 1 & 0  \end{array}\right)$\\
$D^{E''}$ & $\left(\begin{array}{cc} 1 & 0 \\ 0 & 1  \end{array}\right)$ & $\left(\begin{array}{cc} -1 & 0 \\ 0 & -1  \end{array}\right)$ &  $\left(\begin{array}{cc} 0 & -1 \\ 1 & -1  \end{array}\right)$
   &  $\left(\begin{array}{cc} 0 & 1 \\ 1 & 0  \end{array}\right)$\\
\hline
\end{tabular}
\caption{Matrices for irreducible representations of   $D_{3h}$ point group}
\label{table:d7}
\end{table}

The k-vector star  is
\begin{eqnarray}
* {\bf K}=\left\{{\bf K},{\bf K}'\right\}=\left\{\left(\frac{2\pi}{3}, \frac{2\pi}{3\sqrt{3}}\right), \left(\frac{2\pi}{3}, -\frac{2\pi}{3\sqrt{3}}\right)\right\}.
\end{eqnarray}

To find induction matrix of the space group representations we have to decompose $P6/mmm$ relative to $P3/mm$
\begin{eqnarray}
P6/mmm=P3/mm \cup (C_2,{\bf 0})P3/mm.
\end{eqnarray}
The coset representatives of $P6/mmm$ relative to $P3/mm$
are
\begin{eqnarray}
\left\{q_1,q_2\right\}=\left\{(E,{\bf 0}), (C_2,{\bf 0})\right\}.
\end{eqnarray}

To find representations of all
the operators of the space group $(W,{\bf t})$, it is enough to find representations of the operators
\begin{eqnarray}
\label{generator}
(E,{\bf t}), (C_6,{\bf 0}),  (U_2,{\bf 0}), (C_i,{\bf 0}).
\end{eqnarray}
The induction matrix is presented in Table \ref{table:d74}

 \begin{table}
\begin{tabular}{|c|ccccc|c|}
\hline
$(W,w)$ & $q_i$ & $q_i^{-1}$  & $q_i^{-1}(W,{\bf t})$  &  $q_j$ & $q_i^{-1}(W,{\bf t})q_j$ & $M_{ij} \neq 0$ \\
 &  &   &   &  & $=(\tilde{W},\tilde{\bf t})$ & \\
\hline
$(E,{\bf t})$ & $(E,{\bf 0})$ & $(E,0)$  & $(E,{\bf t})$ & $(E,0)$  & $(E,{\bf t})$ & $11$ \\
& $(C_2,{\bf 0})$ & $(C_2,{\bf 0})$ & $(C_2,C_2{\bf t})$  & $(C_2,{\bf 0})$ & $(E,C_2{\bf t})$  &  $22$ \\
\hline
$(C_6,{\bf 0})$ & $(E,{\bf 0})$ & $(E,{\bf 0})$  & $(C_6,{\bf 0})$ & $(C_2,{\bf 0})$  & $(C_3,{\bf 0})$ & $12$ \\
& $(C_2,{\bf 0})$ & $(C_2,{\bf 0})$ & $(C_3,{\bf 0})$  & $(E,{\bf 0})$ & $(C_3,{\bf 0})$  & $ 21$\\
\hline
$(U_2,{\bf 0})$ & $(E,{\bf 0})$ & $(E,{\bf 0})$  & $(U_2,{\bf 0})$ & $(U_2,{\bf 0})$  & $(E,{\bf 0})$ & $12$ \\
& $(C_2,{\bf 0})$ & $(C_2,{\bf 0})$ & $(\sigma_v,{\bf 0})$  & $(E,{\bf 0})$ & $(\sigma_v,{\bf 0})$ &  $21$ \\
\hline
$(C_i,{\bf 0})$ & $(E,{\bf 0})$ & $(E,{\bf 0})$  & $(C_i,{\bf 0})$ & $(C_2,{\bf 0})$  & $(\sigma,{\bf 0})$ & $12$ \\
& $(C_2,{\bf 0})$ & $(C_2,{\bf 0})$ & $(\sigma,{\bf 0})$  & $(E,{\bf 0})$ & $(\sigma,{\bf 0})$  & $ 21$\\
\hline
\end{tabular}
\caption{Induction matrix}
\label{table:d74}
\end{table}
Thus we obtain matrices of the irreps of the space group
\begin{eqnarray}
\label{repr}
&&D^{*K,i}(E,{\bf t})=\left(\begin{array}{c|c}e^{-i{\bf K}\cdot {\bf t}}D^{i}_{3h}(E) & 0 \\
\hline
 0 & e^{i{\bf K}\cdot {\bf t}}D^{i}_{3h}(E)\end{array}\right)\nonumber\\
&&=\left(\begin{array}{c|c}e^{-\frac{2\pi i}{3}(2n_1+n_2)}D^{i}_{3h}(E) & 0 \\
\hline
0 & e^{\frac{2\pi i}{3}(2n_1+n_2)}D^{i}_{3h}(E)\end{array}\right)
\end{eqnarray}
\begin{eqnarray}
\label{repr2}
&&D^{*K,i}(C_6,{\bf 0})=\left(\begin{array}{c|c} 0 & D^{i}_{3h}(C_3)  \\
\hline
D^{i}_{3h}(C_3) & 0 \end{array} \right)\nonumber\\
&&D^{*K,i}(U_2,{\bf 0})=\left(\begin{array}{c|c} 0 & D_{3h}^{i}(E)  \\
\hline
D^{i}_{3h}(\sigma_v) & 0 \end{array} \right)\nonumber\\
&&D^{*K,i}(C_i,{\bf 0})=\left(\begin{array}{c|c} 0 & D^{i}_{3h}(\sigma)  \\
\hline
D^{i}_{3h}(\sigma) & 0 \end{array} \right).
\end{eqnarray}
Representation of any other operator belonging to the space group can be ontained using Eq. (\ref{repr}).

\section{Destruction of Dirac points}

In this note we also apply the group theory to analyze  what happens at the points ${\bf K},{\bf K}'$  in   graphene on a substrate with
a perfectly commensurate superlattice potential, which has the same point symmetry $D_{6h}$ as the  graphene.
We consider explicitly  a
$\sqrt{3}\times\sqrt{3}$  superlattice,
known as the Kekule distortion of the honeycomb lattice
\cite{farjam}.
In this case the allowed translations in the space group are to the vector
\begin{eqnarray}
{\bf t}=\frac{3\tilde{n}_1}{2}\left(1,\sqrt{3}\right)+\frac{3\tilde{n}_2}{2}\left(1,-\sqrt{3}\right),
\end{eqnarray}
which being substituted into Eq. (\ref{n}) gives $2n_1+n_2=3\tilde{n}_1$. Thus the representation
of the translation operator in Eq. (\ref{repr}) becomes
\begin{eqnarray}
\label{repr3}
D^{*K,i}(E)=\left(\begin{array}{c|c}D^{i}_{3h}(E) & 0 \\
\hline
0 & D^{i}_{3h}(E)\end{array}\right).
\end{eqnarray}
In simple terms we may consider the Brillouine zone (BZ) of the superlattice as the folding of the original BZ  \cite{farjam,cheianov}.
The folding leads to the identification of
the corners of the original BZ (${\bf K}$ and ${\bf K}'$)  with the  center  $\tilde{\Gamma}$ of the new BZ.

To study the fate of the Dirac points (at the Fermi level) we should consider $E''$ representation as $D^i_{3h}$. We can forget now about space groups
and decompose (reducible) representation of the group $D_{6h}$ defined by the representations of its operators (\ref{repr2}), (\ref{repr3}) with respect to the irreducible representations of the group. It is convenient to use equation
\begin{eqnarray}
\label{ex}
a_{\alpha}=\frac{1}{g}\sum_G\chi(G)\chi_{\alpha}^*(G),
\end{eqnarray}
which shows how many times a given irreducible representation $\alpha$  is contained in a reducible one \cite{landau}.
In Eq. (\ref{ex}) $n_{\alpha}$ is the dimensionality of the irreducible representation $\alpha$, $g$ is the number of elements in the group, $\chi_{\alpha}(G)$  is the character of an operator  $G$ in the irreducible representation $\alpha$ and $\chi(G)$ is the character of the operator  $G$
in the representation being decomposed.

 For the representation considered, the relevant  characters are $\chi(E)=4$, $\chi(C_3)=-2$, $\chi(C_iC_2)=-4$, $\chi(C_iC_6)=2$.  Hence representation of the group $D_{6h}$  realized by the functions $\psi_{z;{\bf K}}^{A,B}$ and $\psi_{z;{\bf K}'}^{A,B}$ in the reconstructed graphene can be decomposed   as
\begin{eqnarray}
\label{rep7}
R=E_{1g}+E_{2u}.
\end{eqnarray}
We see that due to supercell potential two degenerate Dirac points disappear. At the point $\tilde{\Gamma}$ we have two merging bands
realizing representation $E_{1g}$ and another two merging bands
realizing representation $E_{2u}$. In the vicinity of the point  $\tilde{\Gamma}$ the dispersion law of each pair of bands is represented by a pair of isotropic paraboloids
(touching each other at the point  $\tilde{\Gamma}$).

The folding of the BZ, together with the destruction of previously existing Dirac points, leads to appearance of the new ones. In fact, the new BZ
is still a hexagon, and the same symmetry arguments as  were used for graphene can be used to explain appearence of the Dirac points at the corners of the new BZ ($\tilde{\bf K},\tilde{\bf K}'$).
However, these new Dirac points are situated  deep below or high above the Fermi level and, hence, less manifest themselves than Dirac points of
unreconstructed graphene.

V.U.N. acknowledges  support from National Science Council, Taiwan,
Grant No. 100-2112-M-001-025-MY3.

\end{document}